\documentclass[10pt, conference]{IEEEtran}
\IEEEoverridecommandlockouts
\usepackage{cite}
\usepackage{amsmath,amssymb,amsfonts}
\usepackage{algorithmic}
\usepackage{graphicx}
\usepackage{textcomp}
\usepackage{xcolor}
\usepackage{dsfont}
\PassOptionsToPackage{hyphens}{url}
\usepackage[hidelinks]{hyperref}
\begin{document}




\title{QCE'24 Tutorial: Quantum Annealing -- Emerging Exploration for Database Optimization}

\author{
    \IEEEauthorblockN{Nitin Nayak\IEEEauthorrefmark{1}, Manuel Schönberger\IEEEauthorrefmark{2}, Valter Uotila\IEEEauthorrefmark{3}, Zhengtong Yan\IEEEauthorrefmark{3}, Sven Groppe\IEEEauthorrefmark{1}, Jiaheng Lu\IEEEauthorrefmark{3}, and Wolfgang Mauerer\IEEEauthorrefmark{2}}
    \IEEEauthorblockA{\IEEEauthorrefmark{1}University of L{\"u}beck}
    \IEEEauthorblockA{\IEEEauthorrefmark{2}Technical University of Applied Sciences Regensburg}
    \IEEEauthorblockA{\IEEEauthorrefmark{3}University of Helsinki}
}

\maketitle

\begin{abstract}
Quantum annealing is a meta-heuristic approach tailored to solve combinatorial optimization problems with quantum annealers. In this tutorial, we provide a fundamental and comprehensive introduction to quantum annealing and modern data management systems and show quantum annealing's potential benefits and applications in the realm of database optimization. We demonstrate how to apply quantum annealing for selected database optimization problems, which are critical challenges in many data management platforms. The demonstrations include solving join order optimization problems in relational databases, optimizing sophisticated transaction scheduling, and allocating virtual machines within cloud-based architectures with respect to sustainability metrics. On the one hand, the demonstrations show how to apply quantum annealing on key problems of database management systems (join order selection, transaction scheduling), and on the other hand, they show how quantum annealing can be integrated as a part of larger and dynamic optimization pipelines (virtual machine allocation). The goal of our tutorial is to provide a centralized and condensed source regarding theories and applications of quantum annealing technology for database researchers, practitioners, and everyone who wants to understand how to potentially optimize data management with quantum computing in practice. Besides, we identify the advantages, limitations, and potentials of quantum computing for future database and data management research.

\end{abstract}


\section{Introduction}
 Quadratic Unconstrained Binary Optimization (QUBO) problems are a class of optimization problems formulated as quadratic functions with binary variables. Quantum annealers are designed to solve QUBO problems by mapping them directly onto the quantum hardware. 
Reasons for database (DB) experts to deal with quantum annealing (QA) are:

\textbf{Easy to learn:} The first steps in using QA can be done by just studying QUBO problems and how to solve them on quantum annealers. 
To build upon initial research effectively, a deeper background in quantum computing (QC) is beneficial to understand which formulations of constraints are suitable and which are too complex to run on quantum annealers. For instance, in more intricate problems, the translation of linear and quadratic coefficients from a QUBO problem into qubit bias and coupling values for the quantum annealer does not scale well. 

\textbf{Importance of quantum annealing for DB research:} 
Analyzing survey papers about QC for DBs, the solutions of certain subareas of quantum accelerated DBs are dominated by approaches that can be run on a quantum annealer. Specifically, in the subareas of database query optimization and transaction management, 75\% of the contributions described in \cite{Yuan23QDSM} utilize quantum annealers. 

\textbf{Approximating NP-hard problems:} For some NP-hard problems like the subset sum
and the maximum cut problems, the numbers of variables in the corresponding QUBO formulas scale 
linearly with the problem sizes~\cite{Jiang2023}. The belief that quantum annealers will enjoy an 
increase in performance in the future may be a positive indicator 
of scalability and quantum benefits. However, careful benchmarking
of quantum against classical optimization approaches exposes issues that need to be
addressed, and finding concrete quantum speedups remains a challenge.


\textbf{Tutorial overview:} 
After motivating QA for DB research,
we provide a dive into the basics of QA in Section~\ref{sec:preliminary}. In Section~\ref{sec:use-cases}, we introduce our DB use cases for the QA demo. In the hands-on part, we show how to apply QA to the described DB use cases in Section~\ref{sec:demo}. Finally, we discuss QA opportunities and future work for DB research in Section~\ref{sec:opportunities}. 

\textbf{Related tutorials:} Our previous tutorial in SIGMOD \cite{Winker2023QMLTutorial} has already addressed the benefits of quantum machine learning to data management tasks with a focus on query optimization. However, to the best of our knowledge, no tutorial contains an introduction to QA tailored to DB research. In order to provide a solid basis for future work about QA accelerated DBs, we will introduce the basics of QA and deliver a hands-on tutorial on applying QA to three important data management tasks---join order optimization, transaction scheduling, and virtual machine and task allocation.

\textbf{Contributions:} To the best of our knowledge, this is the first tutorial to discuss QA approaches for DB research. This tutorial helps DB experts get into QA easily (by providing a dense introduction), newbies and practitioners discover the possibilities of QA by showing how to interact with quantum annealers with code examples, and researchers develop new approaches utilizing QA.

\section{Basics of quantum annealing}\label{sec:preliminary}

\textbf{Quantum annealers.} Quantum computing (QC) is a paradigm that consists of multiple computational models that utilize various quantum mechanical phenomena. For example, quantum computers can be implemented using superconducting transmons, trapped ions, trapped neutral atoms, or photonics. Most of the current quantum computers implement the quantum circuit model. Given their capability to implement a certain set of elementary operations, they can approximately implement arbitrary unitary operations (i.e., transformations allowed by quantum mechanics), making them \emph{universal}~\cite{Bernstein_Vazirani_1997}. Alternative approaches based on successive measurements on quantum states, or adiabatic transformations~\cite{farhi2000quantum} are equivalent to the circuit model in terms of the efficiency and computational capabilities~\cite{doi:10.1137/S0097539705447323}.

This tutorial focuses on quantum annealing (QA), which is a specific implementation of adiabatic quantum computing (AQC). Quantum annealing follows similar principles as adiabatic quantum computing. In both cases, the problem is encoded as an objective function called a problem Hamiltonian, which also characterizes the physical properties of the system used to perform the computation. The ground state that minimizes the Hamiltonian (equivalent: the quantum state that minimizes the energy of the system characterized by the Hamiltonian) is chosen so that it delivers the solution to an encoded problem. If we start from a simple initial Hamiltonian, whose ground state we know, and gradually transform it into the final Hamiltonian, the adiabatic theorem~\cite{farhi2000quantum} guarantees that the system remains in the ground state. This process minimizes the final Hamiltonian and encodes a solution to the problem of interest.

The crucial difference between QA and AQC is quantum universality, that is, the ability to execute arbitrary unitary transformations (which is a different notion than classical computational universality). Whereas AQC is universal, QA is strongly suspected to be strictly less expressive \cite{Vinci_Lidar_2017} in terms of executable transformations. This does not mean that annealers could not be useful in practice. One commercial vendor of QAs is D-Wave Systems, offering cloud access to their quantum annealers.


Finding use cases with quantum advantage, that is, performance or other improvements of QC over classical approaches, is crucial. Unfortunately, there is no clear evidence that quantum annealers would provide quantum advantage \cite{Hauke_Katzgraber_Lechner_Nishimori_Oliver_2020, Ronnow_Wang_Job_Boixo_Isakov_Wecker_Martinis_Lidar_Troyer_2014}, although they perform well in specific tasks \cite{Wierzbinski_Falco_Roget_Crimi_2023, Ajagekar_Humble_You_2020}. Since the seminal SAT problem can be expressed as QUBO~\cite{Krueger_2020, Nuesslein_2022}, any relevant DB algorithms can be executed on a quantum annealer (of course, not necessarily efficiently). Expressing a database problem as QUBO is a required step but does not automatically imply computational advantage or even the capability to solve hard variants of specific problems. Assessing performance benefits involves analyzing the spectral properties of the underlying Hamiltonian or employing system-wide experimental designs.

\textbf{Algorithms and software.} Problems for quantum annealers are expressed as Quadratic Unconstrained Binary Optimization (QUBO) problems~\cite{10.3389/fphy.2014.00005}. The physics literature often uses equivalent Ising models with slightly different conventions.

Let $\mathds{B} = \left\{0,1\right\}$ be the set of binary values and $n > 0$. The set $\mathds{B}^n$ denotes the set of all binary bit strings of length $n$. The QUBO problem
comprises minimizing the objective function $f_Q \colon \mathds{B}^n \to \mathds{R}$ defined by
\begin{equation}
    f_Q(x) = xQx^{\dagger} = \sum_{i = 1}^{n}\sum_{j = i}^{n} q_{i,j}x_{i}x_{j},
\end{equation}
where $Q$ is a real-valued symmetric $n \times n$ matrix with elements $q_{i,j}$ for $i,j = 1, \ldots, n$. The variable $x$ is a binary-valued column vector of length $n$, and the variable $x^{\dagger}$ is the corresponding binary-valued row vector.



\section{Quantum annealing for databases}
\label{sec:use-cases}

\subsection{Join order optimization}

As a first use case, we consider the join ordering~(JO) problem, which involves determining (near-)optimal orders for joining database relations during query optimisation. Given its large impact on query execution times, JO is among the most fundamental problems in the database domain. While exhaustive search methods can be applied to determine optimal solutions for small queries, they become infeasible for queries of moderate or large sizes. Yet, large queries are common in real workloads~(e.g., \cite{Neumann.2018}), 
and their challenging nature prompts the evaluation of novel approaches, such as quantum computing, against conventional heuristics. To enable JO on QPUs, a variety of JO-QUBO formulations have recently been proposed~\cite{Schoenberger:2023:sigmod, Nayak2023Bushy, schoenberger:23:qdsm, schoenberger:23:pvldb}, which feature JO models of varying complexity. For this tutorial, we consider the most lightweight encoding, proposed in~\cite{schoenberger:23:pvldb}, to reduce the load on currently imperfect quantum hardware. 


\subsection{Transaction scheduling}

For serializability \cite{books/daglib/0020812}, transactions are in conflict with each other if they contain operations accessing the same data object, where this data object is written at least by one of the transactions. For snapshot isolation \cite{10.1145/568271.223785}, transactions writing to the same data object are conflicting. Concurrently running conflicting transactions causes performance overhead in concurrency control mechanisms, which can be avoided by reordering the transactions called \emph{transaction scheduling} (see \cite{luo2010transaction,10.1145/3588706}). 
We model transaction scheduling as a QUBO problem in \cite{bittner2020avoiding} to be presented as the second use case in this tutorial.

\subsection{Task and virtual machine allocation in cloud} 
The third use case for quantum annealing is to allocate tasks and virtual machines to physical machines concerning multiple constraints \cite{10148143}. The binary variables in this QUBO problem are triples $X_{\mathrm{t}, \mathrm{vm}, \mathrm{pm}}$ where $\mathrm{t}$ is a task id, $\mathrm{vm}$ is a virtual machine id and $\mathrm{pm}$ is a physical machine id. The model returns  $X_{\mathrm{t}, \mathrm{vm}, \mathrm{pm}} = 1$ if the task $\mathrm{t}$ should be allocated to virtual machine $\mathrm{vm}$ which should be allocated to physical machine $\mathrm{pm}$. Multiple hard and soft constraints should be satisfied; for example, allocation should utilize the available resources efficiently, and the carbon footprint should be minimized. This framework is integrated with the cloud data center simulator CloudSim \cite{Calheiros_Ranjan_Beloglazov_De_Rose_Buyya_2011}. 




\subsection{Other DB use cases}
Other database use cases utilize QA for database tasks 
like solving the problem of multiple query optimization~\cite{Trummer2016,Fankhauser_2022} to minimize the cost of running a set of queries. In \emph{Solving Hard Variants of Database Schema Matching on Quantum Computers}~\cite{Fritsch2023}, a textbook approach is demonstrated utilizing quantum cloud services and simulators. Given that QUBOs can be automatically translated from many other mathematical optimization notations~\cite{Lobe_2023}, any database task based on optimization is a possible candidate for annealing.

\section{Quantum annealing for databases}
\label{sec:demo}


In our hands-on part of the tutorial, we demonstrate implementations for solving the discussed database applications with quantum annealing. For each application, we discuss (1) the \emph{preprocessing phase}, which may include, for instance, calculating logarithmic coefficients for JO cost approximation, (2) the \emph{encoding phase}, where we construct respective QUBOs, (3) the \emph{optimisation phase}, where the QUBO problem may be solved on real QA devices, or by simulated annealing on a local device, and finally (4) the \emph{readout phase}, which translates annealing solutions to the original application. 


This tutorial will incorporate three hands-on demonstrations to provide audiences with a tangible and immersive experience regarding the practical implementation of quantum annealing in the context of databases: (1)~We will demonstrate quantum annealing for join ordering, optimising queries joining up to ten relations. We specifically discuss encoding steps and a readout method tailored to join ordering, translating each annealing solution into a valid join order. (2) Transaction scheduling: We will demonstrate transaction scheduling by converting the conflicting transaction problem to a QUBO and solving it using QA and simulated annealing. (3) Task and VM allocation in the cloud: Besides demonstrating the quantum annealing solution for the virtual machine and task allocation problem, we demonstrate how a quantum computing solution can be integrated with the cloud data center using the cloud data center simulator. 

\section{Future work and opportunities}\label{sec:opportunities}

Future works about quantum annealing for DBs include:

1) What database problems could be potentially solved by quantum annealing?  These problems are widespread in database fields, such as join order selection, knob tuning, index tuning, transaction scheduling, and query scheduling. 2) How do we transform and encode the database problems onto the quantum annealer? This involves converting the logical representation of database optimization problems into a form suitable for the quantum hardware. 3) What are the advantages of quantum annealing for databases? The possible benefits may include exploiting the quantum properties of quantum annealing (parallelism, superposition, and entanglement) to outperform the classical algorithms. 4) What are the limitations of quantum annealing for databases?  5) How can hybrid quantum-classical algorithms that leverage the strengths of both quantum and classical computing be developed?  6) How can a quantum annealer be integrated with the traditional database infrastructure and hardware to enable flexible and efficient deployment? 


\section{Tutorial length and intended audience}


\noindent \textbf{Learning outcomes.}
The main learning outcomes of this tutorial are: 
(1)~understanding the impact and remarkable progress of quantum annealing;
(2)~learning basics on quantum annealing, such as QUBO problems and software;
(3)~understanding application cases of quantum annealing for database optimization;
(4)~a take-away message about research challenges and open problems; and 5) three different hands-on demonstrations of quantum annealing for database optimization. Practitioners and students will be able to build an extensive understanding and grasp the latest trends and state-of-the-art techniques in quantum annealing. 

\noindent \textbf{Tutorial length.} The tutorial is scheduled for three hours, comprising an initial one-and-a-half-hour segment dedicated to the background and theoretical aspects of quantum annealing. This will be followed by three practical demonstrations illustrating its applications in databases, occupying the remaining half-hour. This tutorial allows us to delve deep into the realms of quantum computing and quantum annealing, allowing sufficient time to address audience questions and provide a more comprehensive exploration of this subject.

\noindent \textbf{Intended audience.}
This tutorial is intended for a wide audience that may include academic researchers and students as well as industrial developers and practitioners who want to understand the impact of quantum annealing on databases. Basic knowledge of linear algebra is sufficient to follow the tutorial. 
Some background in combinatorial optimization and simulated annealing algorithms would be useful. 

\section*{Tutorial instructors}
\noindent \textbf{Nitin Nayak} is a Ph.D. student researching query optimization using quantum computing (\hspace{1sp}\cite{Nayak2023Bushy, Nayak2024JOOSplit}) at the University of L{\"u}beck.

\noindent \textbf{Manuel Schönberger} is a Ph.D. student researching quantum computing for data management at the Technical University of Applied Sciences Regensburg.

\noindent \textbf{Valter Uotila} is a Ph.D. student researching quantum computing applications for databases and data management at the University of Helsinki. He is also interested in applied category theory \cite{10.1007/978-3-030-93663-1_2,10.14778/3476311.3476314,Uotila_2022}, and its synergy with quantum computing and databases \cite{Uotila_2023}. He has achieved top-3 placements in international quantum computing hackathons, such as QHack 2022 and 2023, as well as in QIA's Quantum Internet Application Challenge and BMW's Quantum Computing for Automotive Challenges.

\noindent \textbf{Zhengtong Yan} is a Ph.D.\ student researching quantum computing and reinforcement learning for databases at the University of Helsinki.

\noindent \textbf{Sven Groppe} is a Professor at the University of L{\"u}beck and the project coordinator of the project QC4DB -- Accelerating Relational Database Management Systems via Quantum Computing. His research includes the integration (\hspace{1sp}\cite{GroppeHM3PDatabase20,Groppe21Semantic}) of quantum accelerators into DBMS (\hspace{1sp}\cite{bittner2020avoiding,OJCC_2020v7i1n01_Bittner,Groppe2021Grover,Groppe22QDM,Gruenwald2023Index,Nayak2023Bushy,Nayak2024JOOSplit,Winker2023QMLTutorial,Winker2023QMLJOO,Calikyilmaz2023Vision,Vogrin2024QGNN}), and high-level quantum programming languages \cite{Hans2022_5}. Sven Groppe's over 190 co-authors of over 180 publications are affiliated with organizations in 28 countries on 6 continents. 
For more details about his academic career, visit \url{https://www.ifis.uni-luebeck.de/~groppe}.

\noindent  \textbf{Jiaheng Lu} is a Professor at the University of Helsinki, Finland.  His current research interests focus on multi-model databases and quantum computing for database applications.  He has written four books on  Hadoop and NoSQL databases, and more than 130 journal and conference papers published in SIGMOD, VLDB, TODS, etc. He was the workshop co-chair of Keyword search and data exploratory with ICDE 2016,
Keyword search on structured data with SIGMOD 2012, 
 Cloud databases with CIKM 2010, and Quantum Data Science and Management (QDSM) with VLDB, 2023 and 2024.

\noindent\textbf{Wolfgang Mauerer} is a Professor of \href{https://www.lfdr.de}{Quantum Computer Science} at the
Technical University of Applied Sciences Regensburg, and a Senior Research Scientist at Siemens Technology. He has published strongly multi-disciplinary work in venues and journals from Nature Photonics and Physical Review Letters via ICSE and TSE to SIGMOD and VLDB.

\bibliographystyle{IEEEtran}
\bibliography{IEEEabrv,ref}

\end{document}